\documentclass[11pt]{article}
\usepackage{graphicx}
\usepackage{amssymb}

\newcommand{\e}{\epsilon}

\newcommand{\eqref}[1]{(\ref{#1})}

\newtheorem{remark}{\textbf{Remark}}[section]

\newtheorem{definition}{\textbf{Definition}}[section]

\begin{document}
\title{\noindent Sparse Time Frequency Representations and Dynamical Systems }

\author{Thomas Y. Hou \thanks{Applied and Comput. Math, MC 9-94, Caltech,
Pasadena, CA 91125. {\it Email: hou@cms.caltech.edu.}}  Zuoqiang Shi \thanks{Mathematical Sciences Center, Tsinghua University, Beijing, China, 100084. {\it Email: zqshi@math.tsinghua.edu.cn.}}  Peyman Tavallali \thanks{Applied and Comput. Math, MC 9-94, Caltech,
Pasadena, CA 91125. {\it Email: ptavalla@caltech.edu.}}}
\maketitle
\begin{abstract}
In this paper, we establish a connection between the recently developed
data-driven time-frequency analysis \cite{HS11,HS13-1} and the classical second order differential equations. 
The main idea of the data-driven time-frequency analysis is to decompose a multiscale signal into a sparsest collection of Intrinsic Mode Functions (IMFs) over the largest possible dictionary  via nonlinear optimization. These IMFs are of the form $a(t) \cos(\theta(t))$ where the amplitude $a(t)$ is positive and slowly varying. The non-decreasing phase function $\theta(t)$ is determined by the data and in general depends on the signal in a nonlinear fashion. One of the main results of this paper is that we show that each IMF can be associated with a solution of a second order ordinary differential equation of the form $\ddot{x}+p(x,t)\dot{x}+q(x,t)=0$. Further, we propose a localized variational formulation for this problem and  develop an effective $l^1$-based optimization method to recover $p(x,t)$ and $q(x,t)$  by looking for a sparse representation of $p$ and $q$ in terms of the polynomial basis. Depending on the form of nonlinearity in $p(x,t)$ and 
$q(x,t)$, we can define the degree of nonlinearity for the associated IMF.
This generalizes a concept recently introduced by Prof. N. E. Huang et al. \cite{Huang11}. Numerical examples will be provided to illustrate the robustness and stability of the proposed method for data with or without noise. This manuscript should be considered as a proof of concept.
\end{abstract}
\maketitle

\section{Introduction}

In many scientific applications such as biology, the underlying physical problem is so complex that we often do not know what is the appropriate governing equation to describe its dynamics. Typically, there are several dominating components that could contribute to the complex phenomena of the underlying physical solution. It is likely that each dominating component can be characterized by a dynamical system. Although we do not know the precise governing equation for these complex phenomena, we can collect a lot of data to characterize the solution of the underlying physical system. A very interesting question to ask is whether or not it is possible to obtain some qualitative understanding of different dominating components from the data that we collect. One of the most important questions is whether the underlying dynamical system is linear or nonlinear. If it is nonlinear, can we quantify the degree of nonlinearity of the underlying dynamical system? In this paper, we attempt to provide one possible approach via a recently proposed data-driven time-frequency analysis method \cite{HS11,HS13-1}.

The most commonly used definition of linearity is that the output of a system is linearly dependent on the input. But this definition 
is not very practical since we may not even know the governing system precisely. It is not easy to define what is input and what is output without knowing the governing system. Another difficulty is that the solution typically consists of several dominating components each of which accounts for a different physical mechanism. Some of these mechanisms may be linear and others may be nonlinear. Thus it is not a good idea to work on the entire data directly. We need to first decompose the data into several dominating components and then try to analyze these components separately. How to extract these intrinsic physical components from the data without compromising their hidden physical structure and integrity is highly nontrivial. For the data that we collect from a nonlinear system, such as the stokes wave, the classical Fourier or wavelet analysis would 
decompose the signal to a collection of fundamental components and harmonics. Each of the components, whether it is a fundamental or harmonic component, looks like a linear signal. A data analysis method based on these linear transformations would suggest that the signal is a superposition of linear components corresponding to a linear system rather than 
a nonlinear system. 

The Empirical Mode Decomposition (EMD) method of Huang et al \cite{Huang98} provides a completely new way to analyze nonlinear and nonstationary signals. The EMD method decomposes a signal into a collection
of intrinsic mode functions (IMFs) sequentially. 
The basic idea behind this approach is the removal of the
local median from a signal by using a sifting process and
a cubic spline interpolation of local extrema. The EMD method has 
found many applications, see e.g.  \cite{Huang07,HW08,WHC09}. One 
important property of these IMFs is that they give physically meaningful 
Hilbert spectral representation. On the other hand, since the EMD 
method relies on the information of local extrema of a signal, it
is unstable to noise perturbation. Recently, an ensemble EMD method (EEMD)
was proposed to make it more stable to noise perturbation \cite{WH09}.
Despite of the tremendous success of EMD and EEMD, there is still lack of a theoretical understanding of this method.
We remark that the recently developed synchrosqueezed wavelet transform
by Daubechies, Lu and Wu \cite{DLW11} is another attempt to provide a
mathematical justification for an EMD like method.

Inspired by EMD/EEMD and the recently developed compressed (compressive)
sensing theory \cite{GN03,CRT06a,Candes-Tao06,Donoho06,BDE09}, 
Hou and Shi have recently introduced a data-driven time-frequency analysis method \cite{HS11,HS13-1}. There are two important ingredients of this method. The
first one is that the basis that is used to decompose the data
is derived from the data rather than determined {\it a priori}.
This explains the name ``data-driven'' in our method. Finding such
nonlinear multiscale basis is an essential ingredient of our method.
In some sense, our problem is more difficult than the compressed
(compressive) sensing problem in which the basis is assumed to be
known {\it a priori}. The second ingredient is to look for the
sparsest decomposition of the signal among the largest possible
dictionary consisting of intrinsic mode functions. In our method, we
reformulate the problem as a nonlinear optimization and find the basis
and the decomposition simultaneously by looking for the sparsest
decomposition among all the possible decompositions.

In this paper, we develop a method to quantify the nonlinearity of the 
IMFs given by the data-driven time-frequency analysis method. 
The main idea is to establish a connection between the 
IMFs and the classical second order differential equations. 
The data-driven time-frequency analysis decomposes a multiscale signal into a sparse collection of IMFs. These IMFs are of the form 
$a(t) \cos(\theta(t))$ where the amplitude $a(t)$ is positive and slowly varying. The non-decreasing phase function $\theta(t)$ is determined by the 
data and is in general nonlinear. One of the main results of this paper 
is that we show that each IMF can be associated with a solution of a 
second order ordinary differential equation of the form
$\ddot{x}+p(x,t)\dot{x}+q(x,t)=f(t)$. We further assume that the 
coefficients $p(x,t)$, $q(x,t)$ and $f(t)$ are slowly varying with respect to $t$. Thus, we can freeze these coefficients locally in time and absorb
the forcing function into $q$. This leads to the reduced autonomous second 
order ODE, i.e. $\ddot{x}+p(x)\dot{x}+q(x)=0$. Further, we can reformulate the second order ODE in a conservative form: $\ddot{x}+\dot{P}(x)+q(x)=0,$
where $\frac{d P(x)}{dx} = p(x)$. We then have the following weak 
formulation of the equation by integrating by parts:
$$
<x,\ddot{\phi}>- <P(x),\dot{\phi}>+ <q(x),\phi>=0,
$$
where $<\cdot,\cdot>$ is the standard inner product, and
$\phi$ is a smooth test function of compact support. 
If $p(x)$ and $q(x)$ have a sparse representation in terms of the 
polynomial basis, then we can represent $P(x)$ and $q(x)$ as follows:
$P(x)=\sum_{k=0}^M p_k x^{k+1}$, $q(x)=\sum_{k=0}^M q_k x^k$
for some integer $M>0$. Then we obtain the following weak formulation:
$$
<x,\ddot{\phi}>-\sum_{k=0}^M p_k <x^{k+1},\dot{\phi}>+\sum_{k=0}^M q_k <x^k,\phi>=0. 
$$
Based on the above weak formulation, we can design a $l^1$-based optimization method to solve for $p_k$ and $q_k$,
$$
(p_k,q_k)=\arg\min_{\alpha_k,\beta_{k}}\gamma \sum_{k=0}^M (|\alpha_k|+|\beta_k|)
+\sum_{i=1}^N\left|<x,\ddot{\phi}_i>-\sum_{k=0}^M \alpha_k <x^{k+1},\dot{\phi}_i>+\sum_{k=0}^M\beta_k <x^k,\phi_i>\right|^2
$$
where $\phi_i$'s are smooth test functions of compact support and
$N$ is the number of the test functions. We will provide some guidance 
how to choose these test functions optimally.

The method described above provides a new way to interpret the hidden
intrinsic information contained in the extracted IMF. Depending on the 
local form of nonlinearity in $p(x,t)$ and $q(x,t)$, we can define the 
degree of nonlinearity for each associated IMF. Moreover, we also recover
accurately the coefficients for the nonlinear terms in $p$ and $q$.
This generalizes a similar concept recently introduced by Prof. N. E. Huang et. al. \cite{Huang11}. Numerical examples will be provided to illustrate the robustness and stability of the proposed method.

The organization of the paper is as follows. In section 2, we give a brief
review of the data-driven time-frequency analysis. Section 3 is devoted
to the connection between IMFs and second order ODEs. We will illustrate
through some examples that solutions of many linear and nonlinear second 
order ODEs have solutions that are essentially IMFs. In section 4, we
introduce two numerical methods to extract the coefficients of the 
second order ODE from a given IMF. Based on the degree of nonlinearity
of the extracted coefficients, we introduce the degree of nonlinearity for
each IMF. This is called \emph{nonlinear degree analysis}. In section 5, 
we demonstrate the effectiveness of the proposed method by a number of 
numerical examples. Some concluding remarks are made in Section 6.

\section{A brief review of the data-drive time-frequency analysis}
\label{section-IMF}
The data-driven time-frequency analysis method is based on finding the sparsest decomposition of a signal by solving a nonlinear optimization problem.
First, we need to construct a large dictionary that can be used to
obtain a sparse decomposition of the signal. 
In our method, the dictionary is chosen to be:
\begin{eqnarray}
  \mathcal{D}=\left\{a\cos\theta:\; a,\theta' \;\mbox{is smoother than}\;
\cos\theta,\; \forall t\in \mathbb{R}, \;\theta'(t)\ge 0 \right\}.
\end{eqnarray}
Let $V(\theta,\lambda)$ be the collection of all the functions that are
smoother than $\cos\theta(t)$. In general, it is most effective to construct $V(\theta,\lambda)$ as an overcomplete Fourier basis given below:
\begin{eqnarray}
\label{2-fold-fourier-2}
V(\theta,\lambda)=\mbox{span}\left\{1, \left(\cos\left(\frac{k\theta}{2L_\theta}\right)\right)_{1\le k\le 2\lambda L_\theta}
,\left(\sin\left(\frac{k\theta}{2L_\theta}\right)\right)_{1\le k\le 2\lambda L_\theta}\right\},
\end{eqnarray}
where $L_\theta=\lfloor\frac{\theta(T)-\theta(0)}{2\pi}\rfloor$,
$\lfloor\mu \rfloor$ is the largest integer less than $\mu$,
 and $\lambda\le 1/2$ is a parameter to control the smoothness of
$V(\theta, \lambda)$.
The dictionary $\mathcal{D}$ then becomes:
\begin{eqnarray}
\label{dic-D}
  \mathcal{D}=\left\{a\cos\theta:\; a\in V(\theta,\lambda),\;\theta'\in
V(\theta,\lambda),\mbox{and}\; \theta'(t)\ge 0, \forall
t\in \mathbb{R} \right\} .
\end{eqnarray}
Each element of the dictionary $\mathcal{D}$ is an IMF with inter-wave frequency modulation. By an IMF with inter-wave frequency modulation, we mean that both the amplitude $a(t)$ and the instantaneous frequency $\theta'(t)$ are less oscillatory than $\cos\theta(t)$. In the case when the
instantaneous frequency $\theta'(t)$ is as oscillatory as $\cos\theta(t)$ or more oscillatory than $\cos\theta(t)$, we call this IMF has intra-wave modulation. The IMFs with intra-wave frequency modulation are not included in this dictionary. We will consider the IMFs with intra-wave frequency modulation in the next section. By saying that a function $f$ is less oscillatory than another function $g$, we mean that $f$ contains fewer high frequency modes than those of $g$ or the high frequency mods of $f$ decay much faster than those of $g$.

Since the dictionary $\mathcal{D}$ is highly redundant,
the decomposition over this dictionary is not unique. We need a criterion
to select the ``best'' one among all possible decompositions. We assume
that the data we consider have an intrinsic sparse structure in the
time-frequency plane in some nonlinear and nonstationary basis. However,
we do not know this basis {\it a priori} and we need to derive (or learn)
this basis from the data. Based on this consideration, we adopt sparsity
as our criterion to choose the best decomposition. This criterion yields
the following nonlinear optimization problem:
 \begin{eqnarray}
\label{opt-inter}
\begin{array}{rcc}\vspace{-2mm}
P_\delta:&\mbox{Minimize}& M\\ \vspace{2mm}
&\scriptstyle (a_k)_{1\le k\le M},(\theta_k)_{1\le k\le M}& \\
&\mbox{Subject to:}&\left\{\begin{array}{l}\|f-\sum_{k=0}^M a_k\cos\theta_k\|_{l^2}\le \delta,\\
a_k\cos\theta_k\in \mathcal{D},\;\quad k=0,\cdots,M,\end{array}\right.
\end{array}
\end{eqnarray}
where $\delta$ depends on the noise level of the signal.

The above optimization problem can be seen as a nonlinear $l^0$ minimization problem.
Thanks to the recent developments of compressed sensing, two types of methods 
have been developed to study this problem. Since we have infinitely 
many elements in the basis (in fact uncountably many), we could not generalize basis pursuit directly to solve our problem. On the other hand, matching pursuit can be generalized. 
However, straightforward generalization of matching pursuit to our
nonlinear optimization problem could be ill-conditioned and
would introduce severe interference among different IMFs. In order to
develop a stable nonlinear optimization method and remove the interference,
 we add an $l^1$ term to regularize the nonlinear least squares problem.
This gives rise to the following algorithm based on a
$l^1$ regularized nonlinear least squares. We begin with $r_0=f,\quad k=0$.

\vspace{0.05in}
\noindent
{\bf Step 1}: Solve the following $l^1$-regularized nonlinear least-square problem $(P_2)$:
\begin{eqnarray}
\label{opt-greedy}
\begin{array}{rcl}\vspace{-2mm}
P_2:\quad (a_k,\theta_k)\in &\mbox{Argmin}&
\gamma \|\widehat{a}\|_{l^1} +
\|r_{k-1}-a\cos\theta\|_{l^2}^2\\
&\scriptstyle a,\theta& \\
&\mbox{Subject to:}& a\in V(\theta,\lambda),\quad \theta'\ge 0, \;\forall t\in \mathbb{R},
\end{array}
\end{eqnarray}
where $\gamma >0$ is a regularization parameter and $\widehat{a}$ is the
representation of $a$ in the overcomplete Fourier basis.

\vspace{0.05in}
\noindent
{\bf Step 2}: Update the residual
$r_{k}=f-\sum_{j=1}^{k}a_{j}\cos\theta_{j}.$

\vspace{0.05in}
\noindent
{\bf Step 3}: If $\|r_{k}\|_{l^2}<\epsilon_0$, stop. Otherwise, set $k=k+1$ and go to Step 1.

\vspace{0.05in}
\noindent
If signals are periodic, we can use the standard Fourier basis to construct $V(\theta,\lambda)$ instead of the
overcomplete Fourier basis. The $l^1$ regularization term is not
needed (i.e. we can set $\gamma=0$) since the standard Fourier basis are
orthogonal to each other. For data with poor samples (i.e. the number of samples is not sufficient to resolve the signal)  or for data with poor scale separation, we would still require the $l^1$ regularization even for periodic data.

One of the main difficulties in solving our $l^1$ regularized
nonlinear least squares problem is that the objective functional
is non-convex since the basis is not known {\it a priori}. We need to
find the basis and the decomposition simultaneously. In \cite{HS13-1},
a Gauss-Newton type method was proposed to solve the $l^1$ regularized
nonlinear least squares problem. 

\subsection{Numerical method for IMFs with intra-wave frequency modulation }

The data-driven time-frequency analysis method described in the previous section is applicable to those signals whose IMFs have only inter-wave modulation but do not have intra-wave frequency modulation. As we will see in Section \ref{IMF-ODE}, the IMF with inter-wave frequency modulation is typically associated with a linear second order ODE, while the IMF with intra-wave frequency modulation is associated with a nonlinear second order ODE.
In order to analyze the nature of nonlinearity in a signal, 
we must consider those IMFs with intra-wave frequency modulation. 
In this section, we describe a modified data-driven time-frequency 
analysis method that is capable of decomposing signals which contain 
IMFs with intra-wave frequency modulation.

For a signal that contains IMFs with intra-wave frequency modulation, they still have a sparse decomposition:
\begin{eqnarray}
  f(t)=\sum_{k=1}^M a_k\cos\theta_k,
\end{eqnarray}
where $a_k$ are smooth amplitude functions. An important difference for data with intra-wave modulation is that their instantaneous frequencies, $\theta'_k$, are no longer in $V(\theta_k,\lambda)$. Typically, the phase function has the form
$\theta_k=\phi_k+\e \cos\left(\omega_k \phi_k\right)$, where 
$\phi_k$ is a smooth function, $\e>0$ is a small number and 
$\omega_k$ is a positive integer. 

An essential difficulty for this type of data is that the instantaneous 
frequency, $\theta'_k$, is as oscillatory as or even more oscillatory than
$\cos\theta_k$. In the method proposed in the previous section, we 
assume that $a_k$ and $\dot{\theta}_k$ are less oscillatory than $\cos\theta_k$.
We use this property to construct the dictionary $\mathcal{D}$. In the 
case when an IMF has strong intra-wave modulation, $\theta'_k$ is as 
oscillatory as $\cos\theta_k$. Thus the method described in the previous section 
would not be able to give a good approximation of $\theta_k$. To 
overcome this difficulty, we introduce a shape function, $s_k$, to replace the cosine function. The idea is to absorb the high frequency intra-wave
modulation into the shape function $s_k$. This will ensure that 
$\theta'_k$ is still less oscillatory than $s_k(\theta_k)$. This idea 
was proposed by Dr. H.-T. Wu in \cite{Wu11}, but he did not provide an 
efficient algorithm to compute such shape function. 

Note that $s_k$ is not 
known {\it a priori} and is adapted to the signal. We need to learn 
$s_k$ from the physical signal. This consideration naturally motivates 
us to modify the construction of the dictionary as follows:
\begin{eqnarray}
  \label{dic-intra}
  \mathcal{M}=\left\{a_ks_k(\theta_k):\quad a_k, \theta'_k\in V(\theta_k,\lambda), \;s_k\; \mbox{is $2\pi$-period function}\right\},
\end{eqnarray}
where $V(\theta,\lambda)$ is defined in (\ref{2-fold-fourier-2}) and $s_k$ is an unknown $2\pi$-periodic `shape function' and is 
adapted to the signal. If we choose $s_k$ to be the cosine function, 
then the new dictionary $\mathcal{M}$ is reduced to the dictionary 
$\mathcal{D}$ that we used previously, i.e. $\mathcal{M}=\mathcal{D}$.

We also use "sparsity" as the criterion to select the decomposition over the redundant dictionary $\mathcal{M}$. 
This would give us the following optimization problem:
 \begin{eqnarray}
\label{opt-ori}
\begin{array}{rcc}\vspace{-2mm}
&\mbox{Minimize}& M\\ \vspace{2mm}
&\scriptstyle (s_k)_{1\le k\le M},(a_k)_{1\le k\le M},(\theta_k)_{1\le k\le M}& \\
&\mbox{Subject to:}&\left\{\begin{array}{l}\|f-\sum_{k=0}^M a_k\cdot s_k(\theta_k)\|_{l^2}\le \delta,\\
a_k\cdot s_k(\theta_k)\in \mathcal{M},\;\quad k=0,\cdots,M,\end{array}\right.
\end{array}
\end{eqnarray}
where $\delta$ depends on the noise level of the signal.

The above optimization problem is much more complicated than (\ref{opt-inter}), since the shape function $s_k$ is also unknown instead of being
determined {\it a priori} as in (\ref{opt-inter}).
In order to simplify this problem, we further assume that the non-zero Fourier coefficients of $s_k$ are confined to a finite number of low frequency modes, i.e. for each $s_k$, there 
exists $N_k\in \mathbb{N}$, such that 
\begin{eqnarray}
  s_k(t)\in \mbox{span}\left\{e^{i j t}, j=-N_k,\cdots,N_k\right\}.
\end{eqnarray}
We further assume that we know how to obtain an estimate for $N_k$ by some method. We call this the low-frequency confinement property of $s_k$. Based on this property of $s_k$, we can represent $s_k$ by its Fourier series,
\begin{eqnarray}
  s_k(t)=\sum_{j=-N_{k}}^{N_k}c_{k,j}e^{ijt}.
\end{eqnarray}
For a given $\theta_k$, we can use this representation and apply the singular value decomposition (SVD) to recover the Fourier coefficients $c_{k,j}$ of each $s_k$. This enables us to obtain the shape function $s_k$.
Once we get an approximation of 
the shape function $s_k$, we can use $s_k$ to update $\theta_k$.
This process continues until it converges. The detail of 
this method will appear in a subsequent paper. In this paper, we will focus on using this generalized data analysis method to perform nonlinearity analysis of multiscale data whose IMFs have intra-wave modulation.

\section{IMFs and Second Order ODEs}
\label{IMF-ODE}
One of the main objectives of this paper is to establish a connection 
between an IMF that we decompose from a multiscale signal and a second 
order ODE. Moreover, we propose an effective method to find such 
second order ODE and study the degree of nonlinearity of the associated 
ODE. For a given IMF of the form $x(t)=a(t)\cos\theta(t)$, it is not
difficult to show that it satisfies the following second order
ordinary differential equation:
\begin{equation}
\ddot{x}+\left(-\frac{\ddot{\theta}}{\dot{\theta}}-2\frac{\dot{a}}{a}\right)\dot{x}
+\left(\dot{\theta}^{2}+\frac{\dot{a}\ddot{\theta}}{a\dot{\theta}}+2\left(\frac{\dot{a}}{a}\right)^{2}-\frac{\ddot{a}}{a}\right)x=0.
\label{ODE-IMF-general}
\end{equation}
Let $p(t)=\left(-\frac{\ddot{\theta}}{\dot{\theta}}-2\frac{\dot{a}}{a}\right)$ and $q(t)=
\left(\dot{\theta}^{2}+\frac{\dot{a}\ddot{\theta}}{a\dot{\theta}}+2\left(\frac{\dot{a}}{a}\right)^{2}-\frac{\ddot{a}}{a}\right)$,
then we get a second order ODE
\begin{equation}
\ddot{x}+p(t)\dot{x}+q(t)x=0.
\label{ODE-IMF-general-linear}
\end{equation}
Note that $p(t)$ and $q(t)$ in general depend on $x(t)$. Thus the above 
ODE may be nonlinear in general. This formal connection does not give 
much information about the nature of the ODE. We will perform further 
analysis to reveal the nature of the associated ODE depending on the
regularity of the amplitude, $a(t)$, and the instantaneous frequency, 
$\theta'(t)$, of a given IMF, $a(t)\cos(\theta(t))$.

\subsection{Connection between Linear Second Order ODEs and IMFs}

Many second order linear differential equations with smooth 
coefficients have solutions that have the form of an IMF, i.e.
$x = a\left(t\right)\cos\theta\left(t\right)$. Moreover, the 
corresponding
amplitude $a\left(t\right)$ and the instantaneous frequency 
$\dot{\theta}(t)$ are smoother than $\cos\theta\left(t\right)$. 
To see this, we consider the following linear second order ODE:
\begin{equation}
\ddot{x}+b(t)\dot{x}+c(t) x=0.\label{eq: 2nd Order Homogeneous ODE standart form}
\end{equation}
It can be also rewritten in the following form:
\begin{equation}
\ddot{v}+Q\left(t\right)v=0,
\label{eq: kenser standard homogeneous form}
\end{equation}
where
\begin{eqnarray}
v=e^{\frac{1}{2}\int_0^{t}b\left(\xi\right)d\xi}x,\quad 
Q\left(t\right)=c\left(t\right)-\frac{1}{4}b^{2}\left(t\right)
-\frac{1}{2}\dot{b}\left(t\right).
\end{eqnarray}
Assume that $Q(t) >0 $ and $Q(t) \gg 1 $.
Using the WKB method \cite{BO99}, 
we can get the asymptotic approximation of $v(t)$,
\begin{eqnarray}
  v(t)\sim c_1\cos\left(\int_0^t \sqrt{Q(\xi)}d\xi\right)+
c_2\sin\left(\int_0^t \sqrt{Q(\xi)}d\xi\right).
\end{eqnarray}
In terms of the original variables, the solution of \eqref{eq: 2nd Order Homogeneous ODE standart form} has the form:
\begin{eqnarray}
  x(t)\sim e^{-\frac{1}{2}\int_0^{t}b\left(\xi\right)d\xi}
\left(c_1\cos\left(\int_0^t \sqrt{Q(\xi)}d\xi\right)+
c_2\sin\left(\int_0^t \sqrt{Q(\xi)}d\xi\right)\right),
\end{eqnarray}
which is essentially an IMF without intra-wave frequency modulation
in which both the amplitude and the instantaneous frequency are
smoother than $\cos\theta(t)$ due to the smoothness of $b$ and $Q$.
 
On the other hand, for those IMFs $a(t)\cos\theta (t)$ that do not 
have intra-wave frequency modulation (meaning that both
$a(t)$ and $\dot{\theta}(t)$ are smoother than $\cos(\theta(t))$),
it is easy to see that the 
coefficients $q$ and $p$ given in \eqref{ODE-IMF-general-linear} are 
smooth functions with respect to $t$. This seems to suggest that 
there is a close connection between oscillatory solutions of a
linear second order ODE with smooth coefficients and IMFs without
intra-wave frequency modulation. 

\subsection{IMFs with Intra-wave Frequency Modulation and Nonlinear ODEs}

For IMFs with intra-wave frequency modulation, the situation is much more complicated. 
In this case, the coefficients $p(t)$ and $q(t)$ that appear in equation \eqref{ODE-IMF-general-linear} are no longer smooth 
since $\dot{\theta}(t)$ is not smooth. As we will demonstrate later,
intra-wave frequency modulation is usually associated with a solution of
a nonlinear ODE. 

Consider a conservative system $\ddot{x}=F\left(x\right)$
where $F\left(x\right)=-\frac{dU\left(x\right)}{dx}$ for some smooth
function $U$.
The total energy of the system is $E=\frac{1}{2}\dot{x}^{2}+U\left(x\right)$.
Assume $E>U\left(x\right)$ for all values of $x$ in $D=\left[x_{0},x_{1}\right]$,
except the end points where $E=U\left(x_{0}\right)=U\left(x_{1}\right)$.
It is obvious that $\dot{x}=0$ only at $x_{0}$ and $x_{1}$. Take
$a=\frac{x_{0}+x_{1}}{2}$, $b=\frac{-x_{0}+x_{1}}{2}$. Consequently,
the range of $\frac{x\left(t\right)-a}{b}$ lies within $\left[-1,1\right].$ If we define
\[
\begin{array}{cclc}
 \theta\left(t\right) & = &\arccos\left(\frac{x\left(t\right)-a}{b}\right), & for \quad \dot{x}<0, \, \\
  \theta\left(t\right) & = & \arccos\left(\frac{-x\left(t\right)+a}{b}\right)+\pi, & for \quad \dot{x}>0.
\end{array}
\]
As a result, we have
\[
\begin{array}{cclc}
\dot{\theta}\left(t\right)&=&\frac{-\dot{x}\left(t\right)}{b\sqrt{1-\left(\frac{x\left(t\right)-a}{b}\right)^{2}}}, & \quad for \quad \dot{x}<0 ,\,\\
\dot{\theta}\left(t\right)&=&\frac{\dot{x}\left(t\right)}{b\sqrt{1-\left(\frac{x\left(t\right)-a}{b}\right)^{2}}}, & \quad for \quad\dot{x}>0,\,
\end{array}
\]
and $\dot{\theta}>0$, if $\frac{x\left(t\right)-a}{b}\neq\pm1$.
Now, if $\frac{x\left(t\right)-a}{b}\rightarrow-1$, then $\dot{\theta}\left(t\right)\rightarrow\sqrt{\frac{\ddot{x}}{b}}$.
Remember that $\ddot{x}>0$ as $\frac{x\left(t\right)-a}{b}\rightarrow-1$.
Similarly, we can show that as
$\frac{x\left(t\right)-a}{b}\rightarrow1$, we have 
$\ddot{x}<0$ and
$\dot{\theta}\left(t\right)\rightarrow\sqrt{\frac{-\ddot{x}}{b}}$.
Therefore, the solution of $\ddot{x}=F\left(x\right)$ can be
represented as $x\left(t\right)=a+b\cos\theta\left(t\right)$, where
$a,b$ are constants and $\theta\left(t\right)\in C^{1}\left(t\right)$,
$\dot{\theta}\left(t\right)>0$. The period $T$ of the oscillation,
using $\dot{\theta}$, can be defined as a real positive number such
that $\int_{0}^{T}\dot{\theta}dt=2\pi$. 

To illustrate this point further, we consider the solution of the 
Duffing equation.
The undamped Duffing equation has the form $\ddot{x}+x+x^{3}=0$.
The energy $E$ of the system is $E=\frac{\dot{x}^{2}}{2}+\frac{x^{2}}{2}+\frac{x^{4}}{4}$.
Obviously, the potential energy is $U(x)=\frac{x^{2}}{2}+\frac{x^{4}}{4}$.
Now, assume that the solution varies within the interval
$\left[-A,A\right]$. Due to symmetry,
we look for a solution of the form $x(t)=A\cos\theta(t)$. 
Substituting this into the energy equation gives 
\[
A^{2}\cos^{2}\theta+A^{2}\dot{\theta}^{2}\sin^{2}\theta+\frac{1}{2}A^{4}\cos^{4}\theta=A^{2}+\frac{A^{4}}{2},
\]
which can be further simplified as
\[
\dot{\theta}^{2}=\frac{A^{2}}{2}\left(1+\cos^{2}\theta\right)+1.
\]
The right hand side of this equation is strictly positive. Since
$\dot{\theta} >0$, we obtain
\[
\dot{\theta}=\sqrt{\frac{A^{2}}{2}\left(1+\cos^{2}\theta\right)+1}.
\]
This shows that the solution of the Duffing equation is an IMF with 
intra-wave frequency modulation. We can see that
the peaks and troughs of the signal coincide with the
maximum of the instantaneous frequency $\dot{\theta}$. 

\section{Nonlinear Degree Analysis}

In this section, we propose a new method to analyze the degree of nonlinearity of the IMFs that we decompose from a multiscale signal. We will present an effective optimization method to construct a second order ODE for each IMF.  Moreover, based on the degree of the nonlinearity of the coefficients associated with the second order ODE, we define the degree of nonlinearity for each IMF.

To begin with, we consider the second order ODE of the following type:
\begin{equation}
\ddot{x}+p(x,t)\dot{x}+q(x,t)= f(t),
\label{eq: Nonlinear Sys Ident model}
\end{equation}
where $p(x,t)$, $q(x,t)$ and $f(t)$ are slowly varying with respect 
to $t$.  For example, in case of the Duffing equation, we have
$p(x)=0,\; q(x)=x+x^{3}$.

Based on this assumption, we can freeze $p(x,t)$, $q(x,t)$, and
$f(t)$ locally in time over a local time interval (a few periods) since they 
vary slowly in time. Thus we can replace the above ODE by the
corresponding autonomous ODE over this local time interval 
and absorb $f$ into $q$ (meaning that we can set $f=0$):
\begin{equation}
\ddot{x}+p(x)\dot{x}+q(x)=0. \label{nonlinear-model-local}
\end{equation}
This approximation reduces the level of difficulty significantly.

\subsection{A Strong Formulation}
In order to determine the autonomous ODE locally, we propose to use polynomials to approximate $p(x)$ and $q(x)$,
\begin{eqnarray}
  \label{poly-ab}
  p(x)=\sum_{k=0}^M p_k x^k,\quad q(x)=\sum_{k=0}^M q_k x^k ,
\end{eqnarray}
where $M$ is the order of polynomials which is given {\it a prior}, $p_k,\; q_k$ are unknown coefficients.

One way to get the coefficients $p_k, \; q_k$ is to substitute 
\eqref{poly-ab} to \eqref{nonlinear-model-local}. This leads to 
\begin{equation}
\ddot{x}+\sum_{k=0}^M p_k (x^k)\dot{x}+\sum_{k=0}^M q_k x^k=0. \label{nonlinear-model-strong}
\end{equation}
Then $p_k,\; q_k$ can be obtained by using a least squares method,
\begin{equation}
(p_k,q_k)=\arg\min_{\alpha_k,\beta_{k}} \|\ddot{x}+\sum_{k=0}^M \alpha_k (x^k) \dot{x}+\sum_{k=0}^M \beta_k x^k\|_2.
\label{ls-strong}
\end{equation}
To study the degree of nonlinearity, we are most interested in the 
highest order terms in $p$ and $q$. Further we assume that the 
coefficients $p_{k}$ and $q_k$ are sparse.
Due the strong correlation between $x$ and $\dot{x}$,
the direct least squares proposed in \eqref{ls-strong} would be 
unstable to noise perturbation. In order to stabilize this optimization 
algorithm, we add a $l^1$ term to regularize the least squares and 
look for the sparsest representation, 
\begin{equation}
(p_k,q_k)=\arg\min_{\alpha_k,\beta_{k}}\gamma \sum_{k=0}^M (|\alpha_k|+|\beta_k|)
+\|\ddot{x}+\sum_{k=0}^M \alpha_k (x^k) \dot{x}+\sum_{k=0}^M\beta_k x^k\|_2^2
,\label{l1-ls-strong}
\end{equation}
where $\gamma$ is a parameter to control the sparsity of the coefficients. In order to capture the leading order term, $\gamma$ is chosen to be 
$O(1)$. In the following examples, $\gamma$ is chosen to be 2.

In the method described above, we need to compute $\ddot{x}$ and $\dot{x}$. This tends to amplify the error introduced in our approximation of the IMF,
$x$. Next, we will introduce another method based on the weak formulation 
of the second order ODE.

\subsection{A Weak Formulation}

In this section, we will introduce a $l^1$-based optimization based on
a weak formulation for the second order ODE. 
Let $P(x)$ be the primitive function of $p(x)$, i.e.
$\dot{P}(x)=p(x)\dot{x}$. Then the ODE can be rewritten in a
conservation form:
\begin{equation}
\ddot{x}+\dot{P}(x)+q(x)=0. \label{nonlinear-model-conserv}
\end{equation}
Suppose the span of time of the signal that we want to study is $[0,T]$.
For any test function $\phi\in C_0^2[0,T]$ satisfying
$\dot{\phi}(0)=\dot{\phi}(T)=0$, we have the following
weak formulation of the equation by performing integration by parts:
\begin{equation}
<x,\ddot{\phi}> - <P(x),\dot{\phi}> + <q(x),\phi> = 0, \label{formulation-weak}
\end{equation}
where $<\cdot,\cdot>$ is the standard inner product.

If $p(x),q(x)$ can be approximated by polynomials as what we have done in \eqref{poly-ab}, 
then $P(x)$ and $q(x)$ can be expanded in terms of polynomial basis:
\begin{eqnarray}
  \label{poly-Q}
  P(x)=\sum_{k=0}^M p_k x^{k+1},\quad q(x)=\sum_{k=0}^M q_k x^k. 
\end{eqnarray}
Then we get 
\begin{equation}
<x,\ddot{\phi}>-\sum_{k=0}^M p_k <x^{k+1},\dot{\phi}>+\sum_{k=0}^M
q_k <x^k,\phi>=0. 
\end{equation}
Using this formulation, we can design the following optimization problem to solve for $p_k$ and $q_k$,
\begin{equation}
(p_k,q_k)=\arg\min_{\alpha_k,\beta_{k}}\gamma \sum_{k=0}^M (|\alpha_k|+|\beta_k|)
+\sum_{i=1}^N\left|<x,\ddot{\phi}_i>-\sum_{k=0}^M \alpha_k <x^{k+1},\dot{\phi}_i>+\sum_{k=0}^M\beta_k <x^k,\phi_i>\right|^2 ,\label{l1-ls-weak}
\end{equation}
where $N$ is the number of the test functions that we use. 
In our computations, we choose $N=2M$ to make sure that we have 
enough measurements to determine the coefficients. The test functions 
that we use are given below:
\begin{eqnarray}
  \phi_i(t)=\left\{\begin{array}{ll}
\frac{1}{2}(1+\cos(\pi (t-t_i)/\lambda)),& -\lambda  <t-t_i<\lambda ,\\
 0,& \mbox{otherwise},
\end{array}\right.\quad i=1,\cdots, N,
\end{eqnarray}
where $t_i$'s ($i=1,\cdots, N$) are the centers of the test functions and the parameter $\lambda$ determines their support. 
In order to enhance stability, we should make the support of the 
test functions as large as possible by choosing a large $\lambda$. 
On the other hand, if the support of $\phi$ is too large, we cannot get 
the high frequency information of the signal, which is essential in
capturing the nonlinearity of the signal. Thus, we should determine 
$\lambda$ based on the balance between stability and resolution. The 
strategy that we use is that to make $\lambda$ as large as possible 
without compromising the resolution.  In our computations, 
$\lambda$ is chosen to be 1/5 of the local period (or wavelength) of 
the signal. After $\lambda$ is determined, we choose $t_i, i=1,\cdots,N$ 
to be uniformly distributed over $[\lambda, T-\lambda]$, where $[0,T]$ is the time span of the signal. 

\begin{remark}
The choice of $\lambda$ depends on the regularity of the signal that
 we want to study. If the signal is nearly singular, we should choose a
small $\lambda$ to make sure that the information of the signal can be 
well captured by the test functions.
\end{remark}

\begin{remark}
If the test functions $\phi_i(t)$ are chosen to be the classical
piecewise linear finite element basis, then the weak formulation 
is equivalent to the strong formulation if we approximate
 $\ddot{x}$ and $\dot{x}$ by a second order central difference
approximation.
\end{remark}

Based on the coefficients that we recover from the signal, we can define 
two indices associate with each IMF to characterize the nonlinearity of 
this IMF.
\begin{definition} (Degrees of Nonlinearity) The degrees of nonlinearity of an IMF are defined to be the following two indices 
\begin{eqnarray}
  I_1=\max\{k: p_k \ne 0,\; k=0,\cdots,M\},\quad   I_2=\max\{k: q_k \ne 0,\; k=0,\cdots,M\} .
\end{eqnarray}
\end{definition}
From the above definition, we can see that the degrees of nonlinearity of the signal correspond to the 
highest order of the nonlinear terms. The case of $I_1=0$ and $I_2=1$  
corresponds to a linear ODE. When $I_1>0$ or $I_2 >1$, the IMF becomes nonlinear. The larger the index is, the more nonlinear the IMF becomes. We not only quantify the degrees of nonlinearity of the IMF, we can also recover the coefficients associated with the leading order nonlinear terms. This information is very helpful in
quantifying how nonlinear an IMF is and may have an important implication
in engineering and biomedical applications.

In practical computations, the signal may be polluted by noise or measurement errors. As a result, our recovery of the coefficients will be 
influenced by these errors. To alleviate this side effect, we set up a 
small threshold $\nu_0$ to enforce sparsity of the coefficients by
keeping only those coefficients that are larger than $\nu_0$. This leads to the following modified
definition of the degrees of nonlinearity:
\begin{eqnarray}
\label{don-practical}
  \quad \quad \quad \quad I_1=\max\{k: |p_k| > \nu_0,\; k=0,\cdots,M\},\quad   I_2=\max\{k: |q_k| > \nu_0,\; k=0,\cdots,M\}
\end{eqnarray} 
In the computations to be presented in the next section, 
we set $\nu_0=0.05$.

The method based on the $l^1$ regularized least squares performs very 
well in identifying those nonlinear terms with large coefficients.
On the other hand, the $l^1$ regularization also compromises the 
accuracy of the coefficients at the expense of producing a sparse
representation of the signal. In order to recover the coefficients 
accurately, we propose the following procedure to improve the
accuracy.
  
First, we identify the dominant coefficients,
\begin{eqnarray}
  \Gamma_1=\{k: |p_k|>\nu_1,\; k=0,\cdots,M\},\quad   
\Gamma_2=\{k: |q_k|>\nu_1,\; k=0,\cdots,M\} .
\end{eqnarray}
In our computations, $\nu_1$ is chosen to be 0.05.

Secondly, we solve a least squares problem without $l^1$ 
regularization to obtain more accurate coefficients for these 
dominant terms,
\begin{equation}
(p_{k_2},q_{k_1})_{k_2\in \Gamma_2, k_1\in \Gamma_1}=\arg\min_{\alpha_{k_2},\beta_{k_1}}\sum_{i=1}^N\left|<x,\ddot{\phi}_i>-
\sum_{k_2\in \Gamma_2} \alpha_{k_2} <x^{k_2+1},\dot{\phi}_i>
+\sum_{k_1\in \Gamma_1}\beta_{k_1} <x^{k_1},\phi_i>\right|^2
,\label{ls-refine}
\end{equation}
\begin{remark}
If the signal is free of noise and accurate, the above refinement 
procedure does help to get more accurate coefficients. But when the 
signal is polluted with noise, the IMF that we extract from the 
signal is not very accurate. In this case, the error of the coefficients
is still relatively large even with the above refinement procedure.
\end{remark}

Before we end this section, we summarize all the discussions to 
give the following algorithm. We first partition the entire 
physical domain into a number of subdomains and localize the
signal locally by multiplying a smooth cut-off function. Then we
apply the above optimization algorithm to the localized signal
to extract the local degrees of nonlinearity of the signal.
\vspace{4mm}

\noindent
{\bf A $l^1$-$l^2$ Refinement Algorithm.}
\begin{itemize}
\item Calculate the phase function $\theta(t)$ of the signal. 
Choose $K$ points $t_j,\; j=1,\cdots,K$ such that the time variation
of $P$ and $q$ is well resolved by the local resolution
$(t_{j+1}-t_j)$.
\item For $j=1:K$

\item \hspace{6mm} Extract the signal around the point $t_j$,
  \begin{eqnarray}
    f_j(t)=f(t)\chi(\theta(t)-\theta(t_j)),\nonumber
  \end{eqnarray}
\hspace{6mm}
where $\chi(t)$ is a cutoff function. In our computations, it is 
chosen to be
\begin{eqnarray}
  \chi(t)=\left\{\begin{array}{ll}
\frac{1}{2}(1+\cos( t/\mu)),& -\mu \pi <t<\mu \pi,\\
 0,& otherwise.
\end{array}\right.\nonumber
\end{eqnarray}
\hspace{6mm}
$\mu$ is a parameter to control the width of the cutoff function.
In this paper, we 

\hspace{6mm}
choose $\mu=3$, which means that for each 
point, we localize the signal within 3 

\hspace{6mm}
periods to perform the 
degrees of nonlinearity analysis.
\item \hspace{6mm} Extract the IMF $c_j$ for $f_j(t)$ using the algorithm in Section \ref{section-IMF}.
\item \hspace{6mm}Solve the optimization problem \eqref{l1-ls-weak} with $x=c_j$ to get the coefficients of the 

\hspace{6mm}polynomials, $P_j(x)$ and $q_j(x)$.
\item \hspace{6mm}(optional) Apply the refinement procedure to update the coefficients.
\item End
\item Calculate the degrees of nonlinearity of the signal according to \eqref{don-practical}.
\end{itemize}

\section{Numerical Results}

In this section, we will show several numerical results to demonstrate the performance of our nonlinearity analysis method 
proposed previously. We first apply our method to study the
degrees of nonlinearity from the signal generated from the solution of
the Van der Pol equation.

 \begin{figure}

    \begin{center}
      \includegraphics[width=0.5\textwidth]{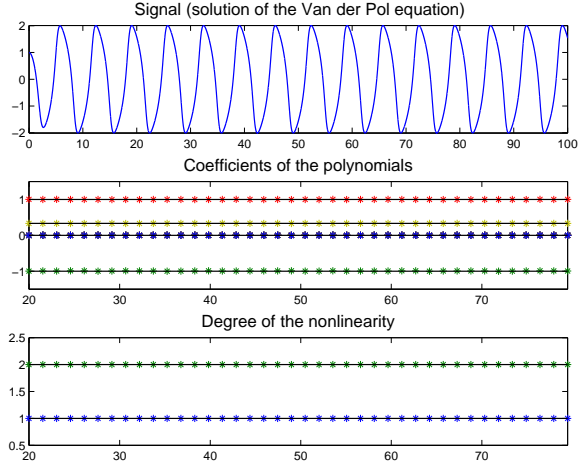}
      \end{center}
    \caption{Top: The solution of the Van der Pol equation; Middle: Coefficients $(q_k,p_k)$ recovered by our method, star points 
$*$ represent the numerical results, black line is the exact one;
Bottom: Nonlinearity of the signal according to the recovered coefficients,
star points 
$*$ represent the numerical results, black line is the exact one.\label{Fig-vanderpol}}
\end{figure}

 \begin{figure}

    \begin{center}
      \includegraphics[width=0.5\textwidth]{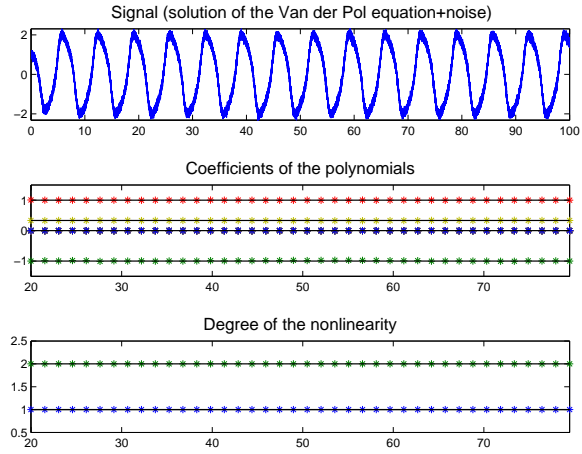}
      \end{center}
    \caption{Top: The solution of the Van der Pol equation
with noise $0.1X(t)$, where $X(t)$ is the white noise with 
standard derivation $\sigma^2=1$; Middle: Coefficients $(q_k,p_k)$ recovered by our method, star points 
$*$ represent the numerical results, black line is the exact one;
Bottom: Nonlinearity of the signal according to the recovered coefficients,
star points 
$*$ represent the numerical results, black line is the exact one.\label{Fig-vanderpol-noise}}
\end{figure}

{\bf Example 1:}
Consider the Van der Pol Equation 
\begin{eqnarray*}
\ddot{x}+\left(x^{2}-1\right)\dot{x}+x=0.
\end{eqnarray*}
The equation is solved from $t=0$ to $t=100$ with the initial condition $x(0)=1,\dot{x}(0)=0$. Fig.
\ref{Fig-vanderpol} shows the original signal and the extracted coefficients and nonlinearity at different times.
We choose $M=10$ in our computations. With this choice,
there are totally 22 coefficients and only three of them are 
not zero. They correspond to $p_1=-1,\; p_2=1/3$ and $q_1=1$ respectively. 
As shown in Fig. \ref{Fig-vanderpol}, we can get almost exact recovery 
of all the coefficients.
When the signal is polluted by noise, our method can still give 
reasonably accurate results, see Fig. \ref{Fig-vanderpol-noise}.

 \begin{figure}

    \begin{center}
      \includegraphics[width=0.5\textwidth]{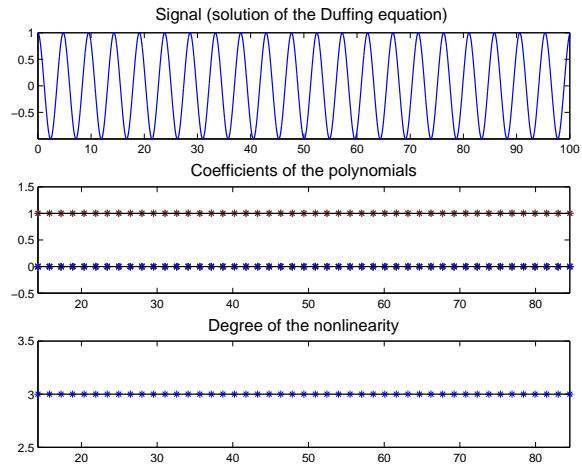}
      \end{center}
    \caption{Top: The solution of the Duffing equation; Middle: Coefficients $(q_k,p_k)$ recovered by our method, star points 
$*$ represent the numerical results, black line is the exact one;
Bottom: Nonlinearity of the signal according to the recovered coefficients,
star points 
$*$ represent the numerical results, black line is the exact one.\label{Fig-duffing}}
\end{figure}

 \begin{figure}

    \begin{center}
      \includegraphics[width=0.5\textwidth]{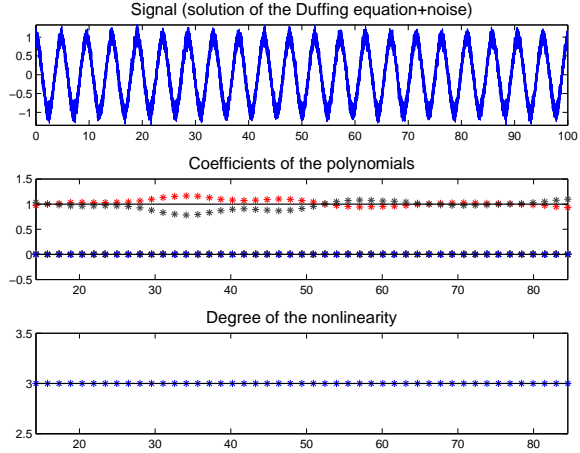}
      \end{center}
    \caption{Top: The solution of the Duffing equation
with noise $0.1X(t)$, where $X(t)$ is the white noise with 
standard derivation $\sigma^2=1$; Middle: Coefficients $(q_k,p_k)$ 
recovered by our method, star points 
$*$ represent the numerical results, black line is the exact one;
Bottom: Nonlinearity of the signal according to the recovered coefficients,
star points 
$*$ represent the numerical results, black line is the exact one.\label{Fig-duffing-noise}}
\end{figure}

{\bf Example 2:}
The second example is the Duffing equation
\begin{eqnarray*}
\ddot{x}+x+x^{3}=0,
\end{eqnarray*}
with initial conditions $x(0)=1$
and $\dot{x}(0)=0$. The solution is also solved from $t=0$ to $t=100$.
Figure \ref{Fig-duffing} shows the original
signal and recovery of the coefficients and degrees of nonlinearity.
Again, we use $M=10$ in our computations. In this case, there are 
actually two coefficients that are not zero, $q_1=1, q_3=1$.

When the signal does not have noise, the recovery is very good for both of the coefficients and the 
degrees of nonlinearity, see Fig. \ref{Fig-duffing}. But when the signal
is polluted by noise, the results for the Duffing equation are not as good as those for the Van der Pol equation, see Fig. \ref{Fig-duffing}.
The reason is that the solution of the Duffing equation is closer to the 
linear sinusoidal wave with $q_3=0$. A small perturbation would 
introduce a large perturbation to the coefficients. 
Nevertheless, even in this case, our method can still give the
 correct degrees of nonlinearity, see Fig.  \ref{Fig-duffing-noise}.

{\bf Example 3:}
The equations in the previous two examples are both autonomous. For this kind of equations, 
the coefficients can be extracted globally, since it does not change over the whole time span.
In order to demonstrate the locality of our method, we consider an 
equation which is not autonomous:
  \begin{eqnarray}
\label{eq-ex3}
\hspace{10mm}    \ddot{x}+a(t)(x^2-1)\dot{x}
+(1-a(t))x^3+x=0
  \end{eqnarray}
where $a(t)=\frac{1}{2}\left(1-\frac{t-100}{\sqrt{(t-100)^2+400}}\right)$. The initial condition is that 
$\dot{x}(0)=0, x(0)=1$ and the equation is solved over $t\in [0,200]$.

\begin{figure}

    \begin{center}
      \includegraphics[width=0.5\textwidth]{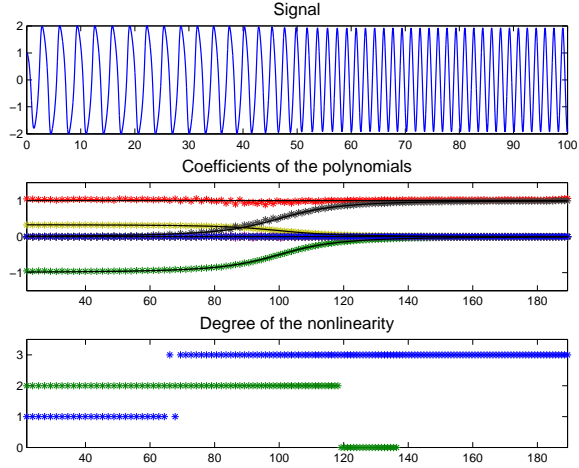}
      \end{center}
    \caption{Top: The solution of the equation given in \eqref{eq-ex3};
 Middle: Coefficients $(q_k,p_k)$ recovered by our method, star points 
$*$ represent the numerical results, black line is the exact one;
Bottom: Nonlinearity of the signal according to the recovered coefficients,
star points 
$*$ represent the numerical results, black line is the exact one.\label{Fig-smooth}}
\end{figure}

\begin{figure}

    \begin{center}
      \includegraphics[width=0.5\textwidth]{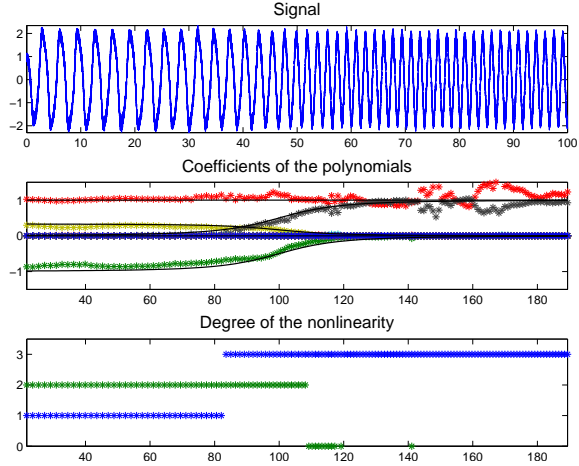}
      \end{center}
    \caption{Top: The solution of the equation given in \eqref{eq-ex3}
with noise $0.1X(t)$, where $X(t)$ is the white noise with 
standard derivation $\sigma^2=1$; Middle: Coefficients $(q_k,p_k)$ recovered by our method, star points 
$*$ represent the numerical results, black line is the exact one;
Bottom: Nonlinearity of the signal according to the recovered coefficients,
star points 
$*$ represent the numerical results, black line is the exact one.\label{Fig-smooth-noise}}
\end{figure}

As we can see, this equation is essentially of the Van der Pol type 
when t is small $(t<100)$. As $t$ increases, the equation
changes to the Duffing type equation gradually. This equation has to 
be analyzed locally. A global approach would predict the wrong degrees
of nonlinearity. We first present our results in Fig. \ref{Fig-smooth} 
when the solution is free of noise. Our method can capture the time
variation of the coefficients very accurately. Even the solution is 
polluted with noise, the results are still with reasonable accuracy, Fig. \ref{Fig-smooth-noise}. 
The error of the coefficients is relatively large when $t>100$.
The reason is that in this region, the equation is qualitatively of 
Duffing type and the Duffing equation is more sensitive
to noise than the Van der Pol equation, as we pointed out in the previous
example.

{\bf Example 4:}
In this example, we consider a more challenging equation, the coefficients 
have a sharp change instead of a smooth transition as in Example 3. 
The equation is given as follows
  \begin{eqnarray}
\label{eq-jump}
    \ddot{x}+\frac{1}{2}\left(1-\mbox{sgn}(t-100)\right)\dot{x}
+\frac{1}{2}\left(1+\mbox{sgn}(t-100)\right)x^3+x=0
  \end{eqnarray}
$\mbox{sgn}(\cdot)$ is the sign function. This equation has 
a sharp transition from the Van der Pol equation to the Duffing equation
at point $t=100$.

When applying our method to analyze the solution of this equation, 
it is not hard to imagine that there would be some problem near the 
transition point, since we require that the coefficients be constants 
over a few periods of the signal. This assumption is not satisfied 
near the transition point.

\begin{figure}

    \begin{center}
      \includegraphics[width=0.5\textwidth]{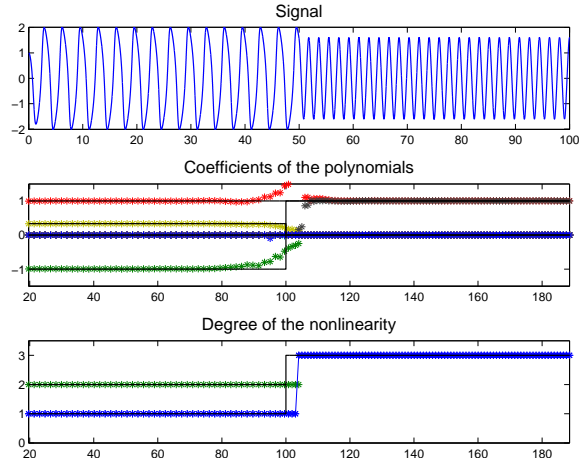}
      \end{center}
    \caption{Top: The solution of the equation given in \eqref{eq-jump};
 Middle: Coefficients $(q_k,p_k)$ recovered by our method, star points 
$*$ represent the numerical results, black line is the exact one;
Bottom: Nonlinearity of the signal according to the recovered coefficients,
star points 
$*$ represent the numerical results, black line is the exact one.\label{Fig-jump}}
\end{figure}

\begin{figure}

    \begin{center}
      \includegraphics[width=0.5\textwidth]{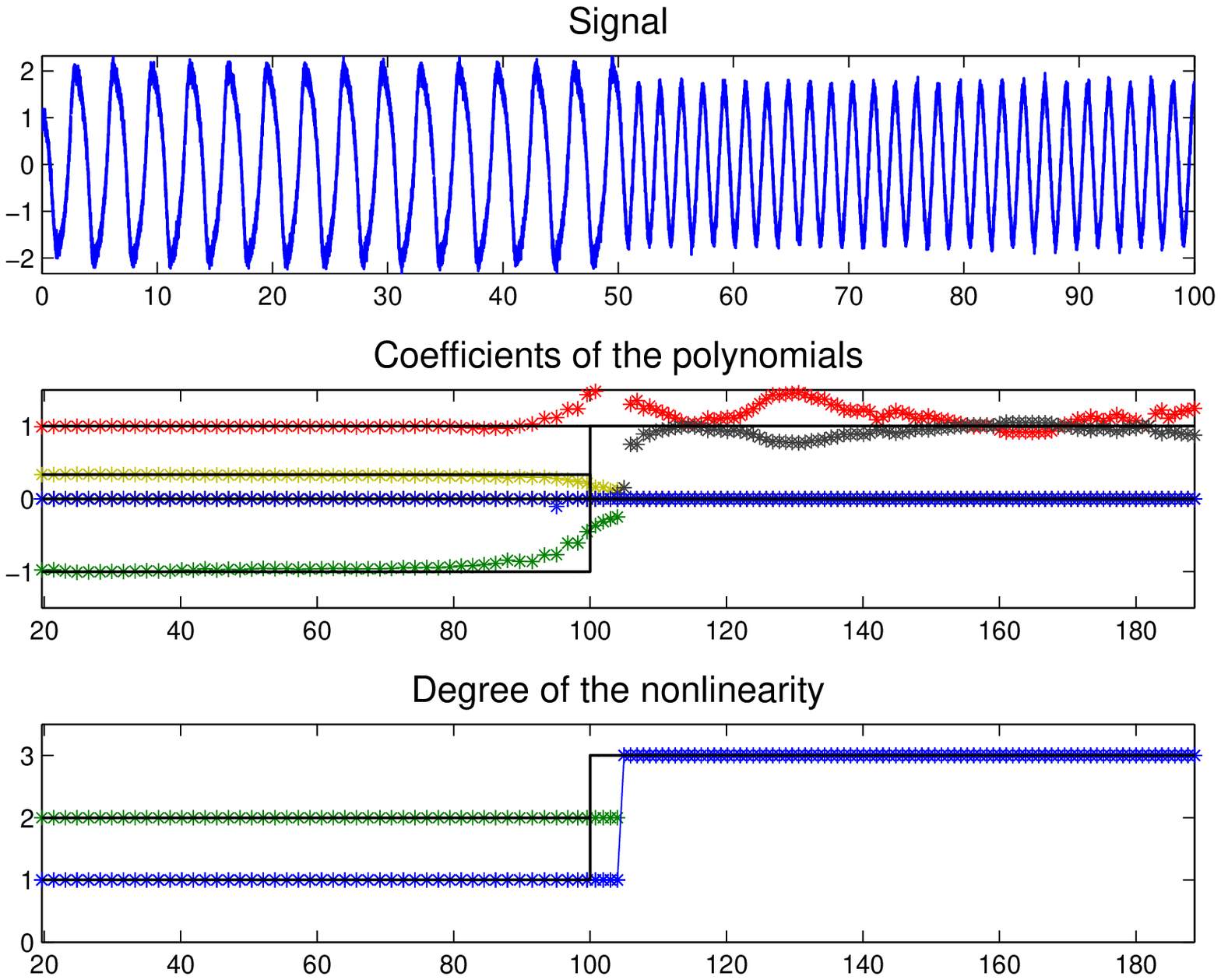}
      \end{center}
    \caption{Top: The solution of the equation given in \eqref{eq-jump}
with noise $0.1X(t)$, where $X(t)$ is the white noise with 
standard derivation $\sigma^2=1$; Middle: Coefficients $(q_k,p_k)$ recovered by our method, star points 
$*$ represent the numerical results, black line is the exact one;
Bottom: Nonlinearity of the signal according to the recovered coefficients,
star points 
$*$ represent the numerical results, black line is the exact one.\label{Fig-jump-noise}}
\end{figure}

We present the results in Fig. \ref{Fig-jump}. It is not surprising
that the error near $t=100$ is very large, but in the region away from 
the transition point, our method still gives a reasonably accurate 
recovery. Due to the poor accuracy near the transition point, our 
method cannot locate the transition point accurately. But the good news 
is that our method does tell us that the nonlinearity of the signal 
changes from the Van der Pol type to the Duffing type, although it 
cannot give the precise location of the transition point. When the
signal is polluted by noise,
the performance of our method is qualitatively the same, see
 Fig. \ref{Fig-jump-noise}.

\begin{figure}

    \begin{center}
      \includegraphics[width=0.5\textwidth]{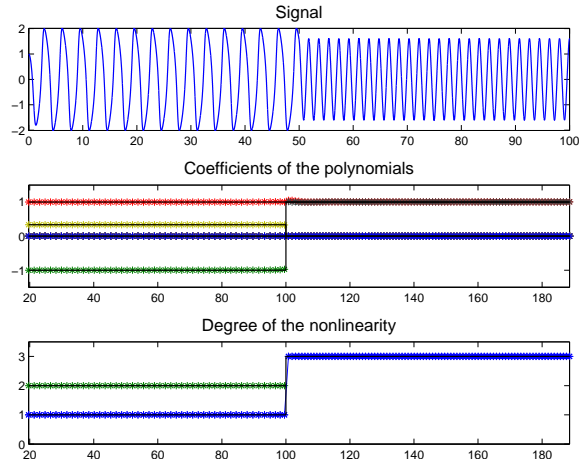}
      \end{center}
    \caption{Top: The solution of the equation given in \eqref{eq-jump}; 
Middle: Coefficients $(q_k,p_k)$ recovered by our method together with 
the trick in ENO method, star points 
$*$ represent the numerical results, black line is the exact one;
Bottom: Nonlinearity of the signal according to the recovered coefficients,
star points 
$*$ represent the numerical results, black line is the exact one.\label{Fig-jump-ENO}}
\end{figure}

\begin{figure}

    \begin{center}
      \includegraphics[width=0.5\textwidth]{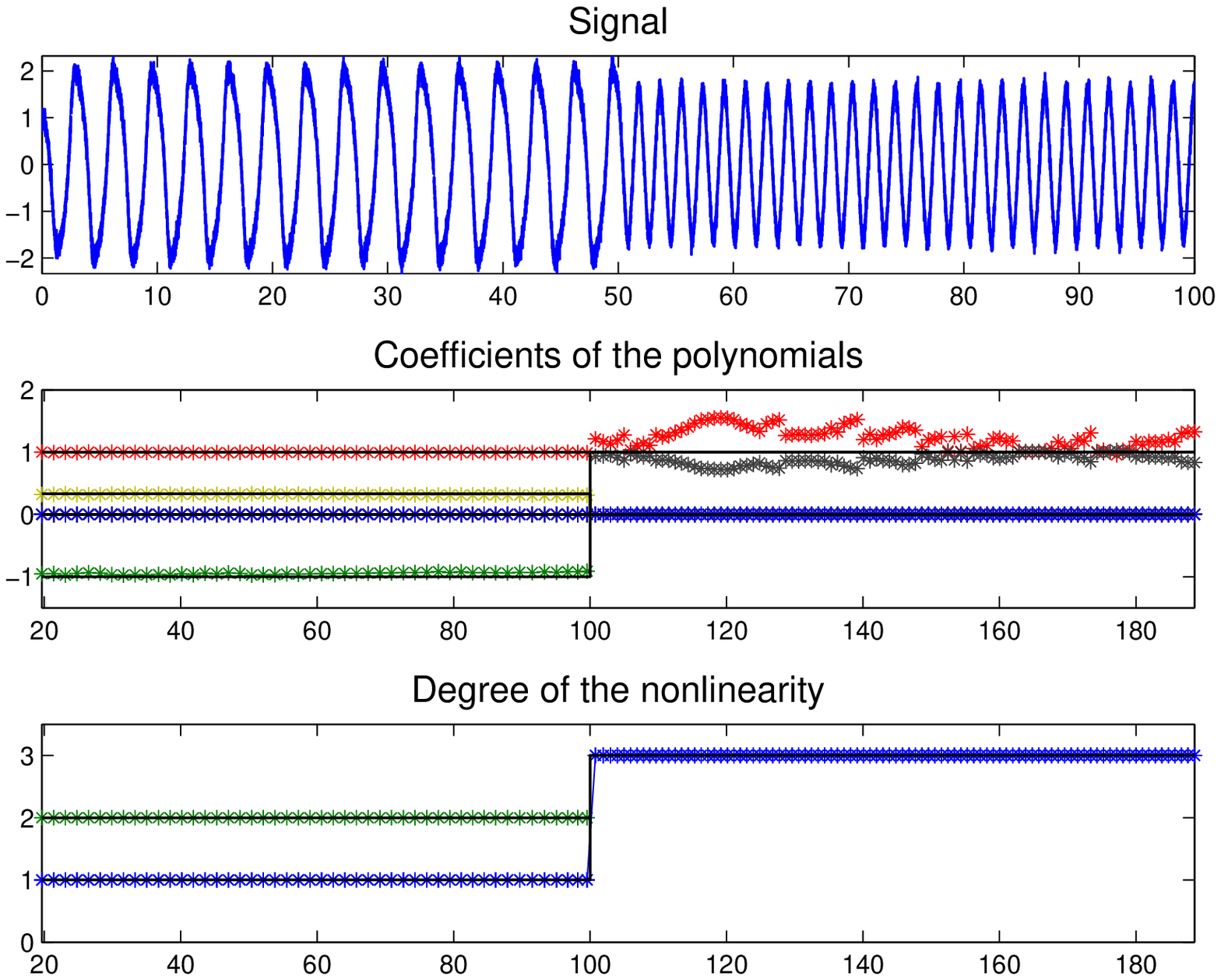}
      \end{center}
    \caption{Top: The solution of the equation given in \eqref{eq-jump} 
with noise $0.1X(t)$, where $X(t)$ is the white noise with 
standard derivation $\sigma^2=1$;
 Middle: Coefficients $(q_k,p_k)$ recovered by our method together with 
the ENO type method, star points 
$*$ represent the numerical results, black line is the exact one;
Bottom: Nonlinearity of the signal according to the recovered coefficients,
star points 
$*$ represent the numerical results, black line is the exact one.\label{Fig-ENO-noise}}
\end{figure}

In order to improve the accuracy in the region near the transition 
point, we combine the idea of the ENO method in computing 
shock waves in fluid dynamics \cite{LeVeque92} with the method that 
we proposed earlier. This gives rise to the following algorithm.  
\begin{itemize}
\item Calculate the phase function $\theta(t)$ of the signal.
Choose $K$ points $t_j,\; j=1,\cdots,K$ such that the time variation 
of $P$ and $q$ is well resolved by the local resolution
$(t_{j+1}-t_j)$.

\item For $j=1:K$
\begin{itemize}
\item[\textbf{S1:}] Extract the signal centered around the point $t_j$ and 
also extract the signal to the left and to the right of $t_j$,
  \begin{eqnarray*}
    f_j^c(t)&=&f(t)\chi^c(\theta(t)-\theta(t_j)),\\
 f_j^l(t)&=&f(t)\chi^l(\theta(t)-\theta(t_j)),\\
f_j^r(t)&=&f(t)\chi^r(\theta(t)-\theta(t_j)),
  \end{eqnarray*}
where $\chi^c(t), \chi^l(t),\chi^r(t)$ are cutoff functions
\begin{eqnarray*}
  \chi^c(t)&=&\left\{\begin{array}{ll}
\frac{1}{2}(1+\cos( t/\mu)),& -\mu \pi <t<\mu \pi,\\
 0,& \mbox{otherwise}.
\end{array}\right.\\
  \chi^l(t)&=&\left\{\begin{array}{ll}
\frac{1}{2}(1+\cos( t/\mu+\pi)),& -2\mu \pi <t<0,\\
 0,& \mbox{otherwise}.
\end{array}\right.\\
  \chi^r(t)&=&\left\{\begin{array}{ll}
\frac{1}{2}(1+\cos( t/\mu-\pi)),& 0 <t<2\mu \pi,\\
 0,& \mbox{otherwise}.
\end{array}\right.
\end{eqnarray*}
As before, we choose $\mu=3$.
\item[\textbf{S2:}] Extract the IMFs $c_j^c, c_j^l, c_j^r$ for $f_j^c(t),f_j^l(t),f_j^r(t)$ respectively.
\item[\textbf{S3:}] Pick up the IMF $c_j^*$ such that the residual $\|c_j^\alpha-f_j^\alpha\|_2$ is minimized over the choices $\alpha=c,l,r$, 
i.e.
\begin{eqnarray*}
  c_j^*=\arg\min_{\alpha\in \{c,l,r\}}\|c_j^\alpha-f_j^\alpha\|_2 .
\end{eqnarray*}
\item[\textbf{S4:}] Solve the optimization problem \eqref{l1-ls-weak} 
with $x=c_j^*$ to get the coefficients of the polynomials, $P_j(x)$ and
$q_j(x)$.
\item[\textbf{S5:}] (optional) Apply the refinement procedure to 
update the coefficients.
\end{itemize}
\item End
\item Calculate the degrees of nonlinearity of the signal according to \eqref{don-practical}.
\end{itemize}
Fig. \ref{Fig-jump-ENO} gives the performance of the above modified
algorithm. The result is much better than the one obtained earlier. 
The coefficients are now accurate over the whole time span of the signal. 
The location of the transition point is also captured accurately.
Even when the signal is polluted with noise, this method is still 
capable of approximating the degrees of nonlinearity and the 
transition point accurately as shown in Fig. \ref{Fig-ENO-noise}.

\begin{figure}

    \begin{center}
      \includegraphics[width=0.5\textwidth]{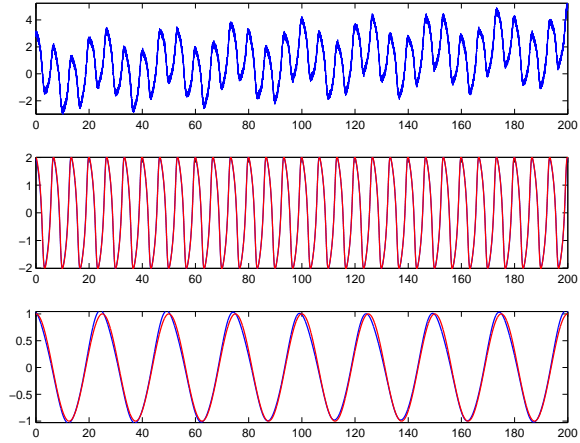}
      \end{center}
    \caption{Top: The signal consists of the solution of the Van der Pol equation and 
a cosine function and a linear trend and noise $0.1X$; 
Middle: The IMF extracted from the signal corresponding to the solution of the Van der Pol equation,
blue: numerical result; red: exact solution;
Bottom: The IMF extracted from the signal corresponding to the cosine function,
blue: numerical result; red: exact solution.\label{Fig-multi-IMF}}
\end{figure}

 \begin{figure}

    \begin{center}
      \includegraphics[width=0.5\textwidth]{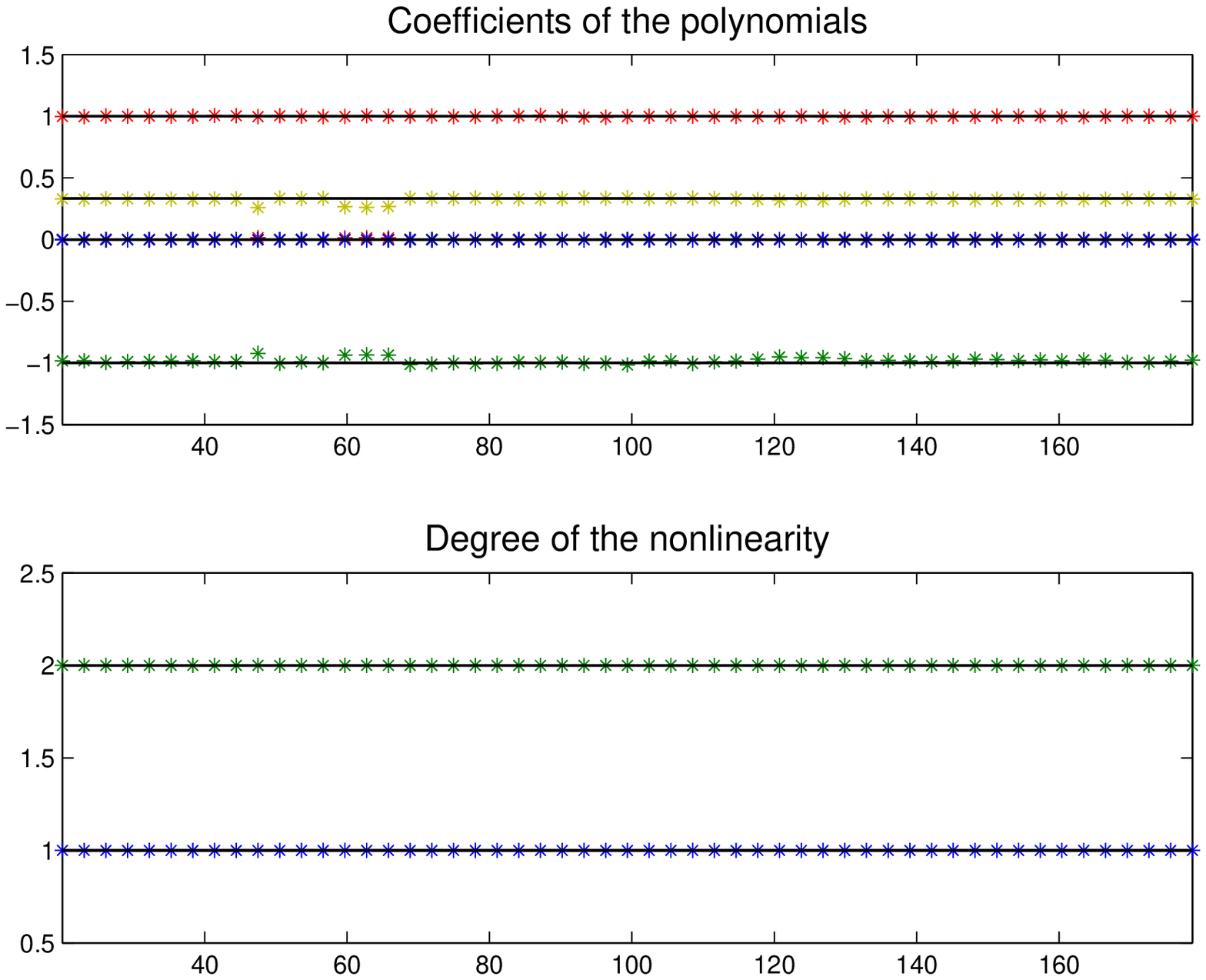}
      \end{center}
    \caption{Top: Coefficients $(q_k,p_k)$ recovered by our method for the 
first IMF in Fig. \ref{Fig-multi-IMF}, star points 
$*$ represent the numerical results, black line is the exact one;
Bottom: Nonlinearity of the signal according to the recovered coefficients,
star points 
$*$ represent the numerical results, black line is the exact one.\label{Fig-multi-coe-1}}
\end{figure}

 \begin{figure}

    \begin{center}
      \includegraphics[width=0.5\textwidth]{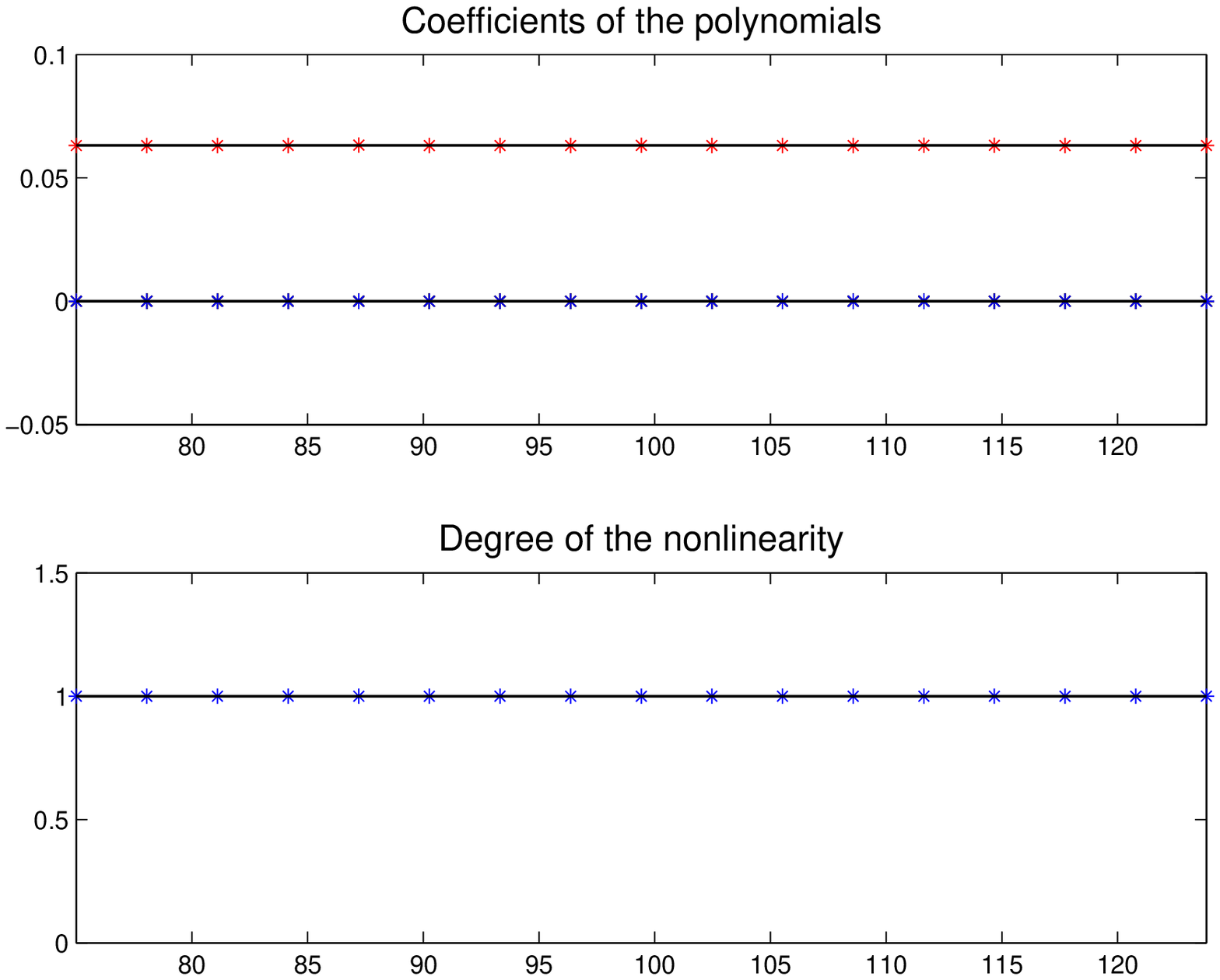}
      \end{center}
    \caption{Top: Coefficients $(q_k,p_k)$ recovered by our method 
for the second IMF in Fig. \ref{Fig-multi-IMF}, 
star points 
$*$ represent the numerical results, black line is the exact one;
Bottom: Nonlinearity of the signal according to the recovered coefficients,
star points 
$*$ represent the numerical results, black line is the exact one.
\label{Fig-multi-coe-2}}
\end{figure}

{\bf Example 5:}
  The signal $f(t)$ we consider in this last example consists of several components,
  \begin{eqnarray}
    f(t)=s(t)+\cos(16\pi t/200)+t/100+0.1X(t),\quad t\in [0,200],
  \end{eqnarray}
where $s(t)$ is the solution of the Van der Pol equation with the initial condition $\dot{x}(0)=0,
x(0)=2$ and $X(t)$ is the white noise with 
standard derivation $\sigma^2=1$. 

For this kind of signal, we have to decompose it to several IMFs first and apply the nonlinearity analysis 
to each IMF to obtain their degrees of nonlinearity. Fig. \ref{Fig-multi-IMF} gives the signal and two IMFs that 
we decompose from the signal. In Fig. \ref{Fig-multi-coe-1} and Fig. \ref{Fig-multi-coe-2}, we present the 
results of the nonlinearity analysis for each IMF. As we can see that for this signal, the performance of our method is still reasonably good.
 
These examples show that our data-driven time-frequency analysis
can be used to detect the type of
nonlinearity (or at least its leading order degree of nonlinearity).
A future goal is to combine this method with statistical 
study to make
the nonlinear system identification algorithm more accurate and 
more stable. 

\section{Concluding Remarks}

In this paper, we have shown that many of the IMFs can be analyzed 
from the point view of dynamical systems. This explains to some extent
why adaptive methods such EMD or our data-driven time-frequency 
analysis method provide a natural way to analyze such signals. 
By establishing a connection between each IMF and a second order
ODE, we can use the information of the associated second order ODE
to obtain further information about the IMF that we extract, including
the degrees of nonlinearity and their energy levels. This information
can be also used to provide a quantitative and qualitative description
of the extracted IMFs of a mutliscale signal. This may prove to be 
very useful in a number of engineering or biomedical applications. 
A possible future direction is to use statistical methods to do
system identification and detect whether the system is linear or 
nonlinear. 

\vspace{0.2in}
\noindent
{\bf Acknowledgments.} We would like to thank Professor Norden E. Huang for a number of stimulating discussions on the topic of the degrees of nonlinearity. This work was supported by NSF FRG Grant DMS-1159138, an AFOSR MURI Grant FA9550-09-1-0613 and a DOE grant DE-FG02-06ER25727.
The research of Dr. Z. Shi was in part supported by a NSFC Grant 11201257.

\end{document}